\documentclass[preprint,12pt,authoryear]{elsarticle}

\usepackage{amsfonts}
\usepackage{amscd}
\usepackage{amssymb}
\usepackage{amsthm}
\usepackage{amsmath}
\usepackage{booktabs}
\usepackage{bm}
\usepackage{caption}
\usepackage{color}
\usepackage{float}
\usepackage{mathrsfs}
\usepackage{mathtools}
\usepackage{multirow}
\usepackage{hyperref}
\usepackage{algorithmic}
\usepackage{algorithm}
\usepackage{subcaption}

\makeatletter
\def\ps@pprintTitle{%
  \let\@oddhead\@empty
  \let\@evenhead\@empty
  \def\@oddfoot{}%
  \let\@evenfoot\@oddfoot}
\makeatother

\begin{document}

\begin{frontmatter}



\title{Privacy-preserving recommender system using \\ the data collaboration analysis for distributed datasets}

\author[label1]{Tomoya Yanagi}
\ead{yanagi.tomoya.ta@alumni.tsukuba.ac.jp}
\author{Shunnosuke Ikeda\corref{cor1}\fnref{label1}}
\ead{ikeda.shunnosuke@gmail.com}
\cortext[cor1]{Corresponding author.}
\author[label2]{Noriyoshi Sukegawa}
\ead{sukegawa@hosei.ac.jp}
\author[label3]{Yuichi Takano}
\ead{ytakano@sk.tsukuba.ac.jp}

\affiliation[label1]{organization={Graduate School of Science and Technology, University of Tsukuba},
            addressline={1--1--1 Tennodai},
            city={Tsukuba-shi},
            postcode={305--8573},
            state={Ibaraki},
            country={Japan}}

\affiliation[label2]{organization={Department of Advanced Sciences, Faculty of Science and Engineering, Hosei University},
            addressline={3--7--2 Kajinocho},
            city={Koganei-shi},
            postcode={184--8584},
            state={Tokyo},
            country={Japan}}

\affiliation[label3]{organization={Institute of Systems and Information Engineering, University of Tsukuba},
            addressline={1--1--1 Tennodai},
            city={Tsukuba-shi},
            postcode={305--8573},
            state={Ibaraki},
            country={Japan}}

\begin{abstract}
In order to provide high-quality recommendations for users, it is desirable to share and integrate multiple datasets held by different parties. 
However, when sharing such distributed datasets, we need to protect personal and confidential information contained in the datasets. 
To this end, we establish a framework for privacy-preserving recommender systems using the data collaboration analysis of distributed datasets. 
Numerical experiments with two public rating datasets demonstrate that our privacy-preserving method for rating prediction can improve the prediction accuracy for distributed datasets. 
This study opens up new possibilities for privacy-preserving techniques in recommender systems. 
\end{abstract}


\end{frontmatter}


\section{Introduction}\label{sec:intro}
\subsection{Background}\label{sec:background}
In recent years, advances in information and communication technology have made it possible for individuals and organizations to access vast amounts of information on a daily basis. 
Against this background, recommender systems have become one of the most successful technologies based on data analytics~\citep{resnick1997recommender,aggarwal2016recommender}.
These systems involve suggesting a personalized list of appealing items based on the users' past preferences.
Various algorithms, including collaborative filtering \citep{resnick1994grouplens} and matrix factorization \citep{koren2009matrix}, have been developed to provide high-quality recommendations for users.
Additionally, these algorithms are actively implemented in a variety of web services~\citep{lu2015recommender,del2021ai}. 

A number of prior studies have focused on improving the prediction accuracy of recommender algorithms~\citep{bobadilla2013recommender}. 
In particular, deep learning techniques have recently received attention and achieved remarkable success in various real-world services~\citep{zhang2019deep,gao2022graph}. 
The recommendation accuracy can be enhanced not only through sophisticated algorithms but also through data fusion, which involves merging multiple datasets into a single, consistent, and clean representation~\citep{bleiholder2009data}. 
Therefore, to develop highly accurate recommender algorithms, it is desirable to share and integrate multiple datasets possessed by different parties.

When sharing such distributed datasets, however, we need to protect personal and confidential information contained in the datasets~\citep{jeckmans2013privacy,himeur2022latest,ogunseyi2023systematic}. 
Privacy is an important issue for recommender systems, which contain information on a large number of registered users~\citep{bobadilla2013recommender}. 
In fact, recommender systems may access users' sensitive information, such as gender, age, and location, to improve the prediction accuracy \citep{sun2018conversational}. 
Therefore, integrating datasets for better recommendations requires an algorithmic framework to protect user privacy. 

\subsection{Related work}\label{sec:related work}
Various privacy-preserving techniques have been used to protect personal and confidential information in recommender systems~\citep{jeckmans2013privacy,himeur2022latest,ogunseyi2023systematic}; these include anonymization, randomization, cryptgraphy techniques, differential privacy, and federated learning. 

Anonymization involves removing personally identifiable information from data, 
whereas randomization aims at modifying data by adding some random noise.  
Although these techniques are readily available in recommender systems~\citep{weinsberg2012blurme,polatidis2017privacy,wei2018improving,saleem2021parking}, they can lead to the loss of key information that is useful in generating accurate recommendations~\citep{ogunseyi2023systematic}. 

Cryptography techniques allow us to conduct data analysis while keeping data secret. 
\citet{canny2002collaborative} was probably the first to apply cryptography techniques to collaborative filtering. 
\citet{nikolaenko2013privacy} designed a recommender system based on matrix factorization using a cryptography technique known as garbled circuits. 
Although various cryptography techniques have been used for privacy protection in recommender systems, these techniques incur substantial computational and communication costs, making them impractical~\citep{ogunseyi2023systematic}. 

Differential privacy is a rigorous mathematical definition of privacy to guarantee that no individual-level information in a dataset is leaked. 
This is typically achieved by adding noise to individual data, where the amount of noise depends on the required level of privacy protection. 
\cite{mcsherry2009differentially} proposed a movie recommendation algorithm that mitigates the impact of noise added for differential privacy through post-processing. 
Various algorithms have been studied to apply differential privacy to collaborative filtering~\citep{aimeur2008alambic,yin2020improved}, matrix factorization~\citep{berlioz2015applying,liu2015fast,ran2022differentially}, and variational autoencoders~\citep{fang2022differentially}. 
However, when a high degree of privacy protection is required, the differential privacy significantly reduces recommendation accuracy~\citep{bagdasaryan2019differential}.

Federated learning aims at training a machine learning model from multiple local datasets while keeping them decentralized. 
The basic strategy consists of training local models from each local dataset and updating the global model by centralizing only the trained parameters.
\cite{ammad2019federated} proposed a federated collaborative filtering algorithm based on matrix factorization. 
Since then, various recommender systems based on federated learning have been proposed for privacy protection~\citep{yang2020federated}; these systems use matrix factorization~\citep{liang2021fedrec++,zhang2023lightfr}, deep neural networks~\citep{liu2022federated,wang2022fast}, and variational autoencoders~\citep{imran2023refrs}. 
However, one of the challenges facing federated learning is the substantial communication costs required to train a machine learning model, which can result in long execution times~\citep{zhang2021survey}.

To overcome these challenges associated with the privacy-preserving techniques mentioned above, \cite{imakura2020data} proposed the data collaboration analysis for distributed datasets. 
This method enables collaborative data analysis by sharing intermediate representations, each being individually constructed for privacy protection by each party from original datasets. 
\cite{bogdanova2020federated} demonstrated that when the number of involved parties is small, the data collaboration analysis consistently outperforms federated learning with lower computational and communication costs. 
\cite{imakura2021accuracy} proved that original datasets can be protected by the data collaboration analysis against insider and external attacks. 
\cite{imakura2023non} proposed an improved version of the data collaboration analysis sharing intermediate representations that are not readily identifiable to the original datasets. 

It is, however, impossible to directly apply the data collaboration analysis to recommender systems. 
A main reason for this is that the data collaboration analysis is specifically designed for regression and classification tasks. 
Therefore, this analysis cannot be performed to impute missing values in the user--item rating matrix used for rating prediction.
To our knowledge, no prior studies have applied the data collaboration analysis to missing value imputation. 

\subsection{Our contribution}\label{sec:contribution}
The goal of this paper is to establish a framework for privacy-preserving recommender systems using the data collaboration analysis. 
For this purpose, we take advantage of the flattened data format used in the factorization machines~\citep{rendle2010factorization}. 
Specifically, we convert a user–item rating matrix into the flattened format with the aim of treating the missing value imputation as the regression analysis. 
This conversion makes it possible to apply the data collaboration analysis to rating prediction for recommender systems. 
Additionally, our algorithm can handle both horizontal and vertical integration of rating matrices.

To verify the effectiveness of our privacy-preserving method for rating prediction, we performed numerical experiments with two public rating datasets. 
For comparison, we implemented two alternative methods: the individual analysis, which uses distributed datasets separately for rating prediction; and the centralized analysis, which merges distributed datasets without preserving privacy.
Numerical results demonstrate that, similar to the centralized analysis, our method using the data collaboration analysis significantly outperformed the individual analysis. 
Moreover, our method improved its prediction accuracy as the number of involved parties increased.

\section{Data collaboration analysis}\label{sec:dca}
In this section, we provide an overview of the data collaboration analysis based on the literature~\citep{imakura2020data,imakura2021collaborative}.
Throughout this paper, we denote the set of consecutive
integers ranging from 1 to $n$ as $[n] \coloneqq \{1, 2, \ldots, n\}$.

\subsection{Distributed datasets}\label{sec:dd}
We suppose that there are $m$ parties, and each party $k \in [m]$ holds the following dataset containing $n(k)$ instances: 
\begin{align}
    \bm{X}^{(k)} 
    \coloneqq \begin{pmatrix}
        \bm{x}^{(k)}_1 \\
        \bm{x}^{(k)}_2 \\
        \vdots \\
        \bm{x}^{(k)}_{n(k)}
    \end{pmatrix} \in \mathbb{R}^{n(k) \times p}, \quad
    \bm{y}^{(k)} 
    \coloneqq \begin{pmatrix}
        y^{(k)}_1 \\
        y^{(k)}_2 \\
        \vdots \\
        y^{(k)}_{n(k)}
    \end{pmatrix} \in \mathbb{R}^{n(k) \times 1}, \label{eq:data_party}
\end{align}
where for each instance $i \in [n(k)]$, $\bm{x}^{(k)}_i \in \mathbb{R}^{1 \times p}$ is a row vector composed of $p$ predictor variables, and $y^{(k)}_i \in \mathbb{R}$ is a response variable to be predicted. 

In the \emph{individual data analysis}, each party $k \in [m]$ uses the dataset (Eq.~\eqref{eq:data_party}) separately to train its own machine learning model. 
In this case, the prediction accuracy of trained models tends to be lower when the size of each dataset is small. 
In the \emph{centralized data analysis}, the datasets held by parties $k \in [m]$ are merged as 
\begin{align}
\bm{X} \coloneqq \begin{pmatrix}
\bm{X}^{(1)} \\
\bm{X}^{(2)} \\
\vdots \\
\bm{X}^{(m)}
\end{pmatrix} \in \mathbb{R}^{n \times p}, \quad 
\bm{y} \coloneqq \begin{pmatrix}
\bm{y}^{(1)} \\
\bm{y}^{(2)} \\
\vdots \\
\bm{y}^{(m)}
\end{pmatrix} \in \mathbb{R}^{n \times 1}, \label{eq:data_all}
\end{align}
where $n \coloneqq \sum_{k=1}^m n(k)$. 
This merged dataset is then used to train a machine learning model; however, sharing datasets is often impossible due to privacy issues. 

\subsection{Intermediate representations}\label{sec:ir}
We aim to analyze multiple datasets collaboratively while keeping them decentralized for privacy protection. 
For this purpose, we use the \emph{anchor dataset}, which is an artificially prepared dataset containing $r$ instances:
\begin{equation}
\bm{S} \coloneqq \begin{pmatrix}
\bm{s}_1 \\
\bm{s}_2 \\
\vdots \\
\bm{s}_r
\end{pmatrix} \in \mathbb{R}^{r \times p}, \label{eq:data_anchor}
\end{equation}
where $\bm{s}_i \in \mathbb{R}^{1 \times p}$ is a row vector corresponding to $p$ predictor variables for each instance $i \in [r]$. 
This dataset can be generated using random numbers or through more sophisticated methods as proposed by \cite{takahashi2021decentralized} and \cite{imakura2023another}.
The anchor dataset is shared by all parties. 

To preserve the privacy of the original datasets (Eq.~\eqref{eq:data_party}), each party $k \in [m]$ applies an individual \emph{encoding function}
\begin{equation}
    f_k:~\mathbb{R}^{1 \times p} \to \mathbb{R}^{1 \times \tilde{p}(k)} \label{eq:enc_fun}
\end{equation}
to all instances (i.e., row vectors) of the original and anchor datasets, $\bm{X}^{(k)}$ and $\bm{S}$, thereby transforming them into the following \emph{intermediate representations}: 
\begin{align}
    \tilde{\bm{X}}^{(k)} 
    \in \mathbb{R}^{n(k) \times \tilde{p}(k)}, \quad 
    \tilde{\bm{S}}^{(k)} 
    \in \mathbb{R}^{r \times \tilde{p}(k)}. \label{eq:data_intrep}
\end{align}
For example, each party can employ dimensionality reduction techniques, such as the principal component analysis and the singular value decomposition, as an encoding function (Eq.~\eqref{eq:enc_fun}). 
Each party $k \in [m]$ sends the intermediate representations (Eq.~\eqref{eq:data_intrep}) and the response vector $\bm{y}^{(k)}$ to the analyzer for data collaboration. 
The privacy of the original datasets is preserved by
not sharing the encoding function with other parties or the analyzer, and 
not sharing the anchor dataset with the analyzer.

\subsection{Collaboration representations}\label{sec:cr}
It is ineffective to merge and analyze the intermediate representations $\tilde{\bm{X}}^{(k)}$ collected from parties $k \in [m]$, because they are created by each party using different encoding functions to preserve privacy. 
To remedy this situation, the analyzer applies the \emph{integration function}  
\begin{equation}
    g_k:~\mathbb{R}^{1 \times \tilde{p}(k)} \to \mathbb{R}^{1 \times \hat{p}} \label{eq:int_fun}
\end{equation}
to all instances (i.e., row vectors) of the intermediate representations (Eq.~\eqref{eq:data_intrep}), resulting in the following $\hat{p}$-dimensional \emph{collaboration representations}:
\begin{align}
    \hat{\bm{X}}^{(k)} 
    \in \mathbb{R}^{n(k) \times \hat{p}}, \quad
    \hat{\bm{S}}^{(k)} 
    \in \mathbb{R}^{r \times \hat{p}}. \notag
\end{align}

\cite{imakura2020data} considered the following linear integration function:
\begin{equation}
g_k(\tilde{\bm{x}}) = \tilde{\bm{x}}\bm{G}^{(k)}, \label{eq:lin_int}
\end{equation}
where $\bm{G}^{(k)} \in \mathbb{R}^{\tilde{p}(k) \times \hat{p}}$ is the coefficient matrix for each party $k \in [m]$. 
In this case, the collaboration representations for party $k \in [m]$ are expressed as follows:
\begin{align}
    \hat{\bm{X}}^{(k)} = \tilde{\bm{X}}^{(k)} \bm{G}^{(k)}, \quad 
    \hat{\bm{S}}^{(k)} = \tilde{\bm{S}}^{(k)} \bm{G}^{(k)}. \label{eq:data_clbrep}
\end{align}
Recall here that the anchor dataset is common among all parties. 
Therefore, the corresponding collaboration representations should be close to each other to ensure data consistency between different parties $k,k' \in [m]$: 
\[
\tilde{\bm{S}}^{(k)} \bm{G}^{(k)} \approx \tilde{\bm{S}}^{(k')} \bm{G}^{(k')} \quad (k \neq k'). 
\]

To estimate linear integration functions (Eq.~\eqref{eq:lin_int}), \cite{imakura2020data} proposed solving a minimum perturbation problem.
This method first computes the singular value decomposition of the following matrix, which consists of the intermediate representations (Eq.~\eqref{eq:data_intrep}) of the anchor dataset: 
\[
\bm{W}_{\tilde{S}} \coloneqq \left(\tilde{\bm{S}}^{(1)} \; \tilde{\bm{S}}^{(2)} \; \cdots \; \tilde{\bm{S}}^{(m)} \right) \in \mathbb{R}^{r \times \tilde{p}},
\]
where $\tilde{p} \coloneqq \sum_{k=1}^m \tilde{p}(k)$. 
Let $\bm{Z} \in \mathbb{R}^{r \times \hat{p}}$ be a target matrix consisting of the left-singular vectors corresponding to the largest $\hat{p}$ singular values of $\bm{W}_{\tilde{S}}$. 

We next calculate the coefficient matrix $\bm{G}^{(k)}$ such that the squared distance between the target matrix $\bm{Z}$ and the collaboration representation (Eq.~\eqref{eq:data_clbrep}) of the anchor dataset will be minimized for each party $k \in [m]$ as follows:
\begin{equation}\label{obj:prob0}
    \min_{\bm{G}^{(k)}} \| \bm{Z} - \tilde{\bm{S}}^{(k)} \bm{G}^{(k)} \|^2_\mathrm{F}.
\end{equation}
We obtain $\bm{G}^{(k)} = (\tilde{\bm{S}}^{(k)})^\dagger \bm{Z}$ as the analytical solution to problem (Eq.~\eqref{obj:prob0}), where $(\tilde{\bm{S}}^{(k)})^\dagger$ is the Moore--Penrose pseudoinverse of the matrix $\tilde{\bm{S}}^{(k)}$.
The generalized eigenvalue problem~\citep{kawakami2024new} and the matrix manifold optimal computation~\citep{nosaka2023creating} were also proposed to estimate linear integration functions (Eq.~\eqref{eq:lin_int}). 

\subsection{Collaborative machine learning}\label{sec:mr}
The collaboration representations (Eq.~\eqref{eq:data_clbrep}) obtained from the integration functions (Eq.~\eqref{eq:lin_int}) enable collaborative machine learning. 
Specifically, after combining the collaboration representations from all parties $k \in [m]$ into a single dataset $\{(\hat{\bm{x}}_i, y_i) \mid i \in [n]\}$, we train a machine learning model
\begin{equation}
h:~\mathbb{R}^{1 \times \hat{p}} \to \mathbb{R} \quad \mbox{such that} \quad h(\hat{\bm{x}}_i) \approx y_i \quad (i \in [n]). \label{eq:ML_model}
\end{equation}
The obtained machine learning model (Eq.~\eqref{eq:ML_model}) and the integration function (Eq.~\eqref{eq:lin_int}) are returned to each party. 
Each party $k \in [m]$ can then acquire a highly accurate machine learning model for the original dataset by adding its encoding function (Eq.~\eqref{eq:enc_fun}) as follows: 
\begin{equation}
h \circ g_k \circ f_k:~\mathbb{R}^{1 \times p} \to \mathbb{R}. \label{eq:fin_model}
\end{equation}

\section{Rating prediction using the data collaboration analysis} \label{sec:dca_rec}
In this section, we first describe a rating prediction problem for recommendations and its data format conversion. 
We then formulate our rating prediction algorithm using the data collaboration analysis for privacy protection. 

\subsection{Data format conversion}
The process of predicting ratings in recommendation systems deals with a user--item rating matrix as shown in Table~\ref{tab:matrixbefore}, where each entry indicates the user's rating for a particular item. 
For example, these ratings can be expressed as a five-point scale of the degree of preferences, or as a binary scale indicating ``likes/dislikes.'' 
We now focus on the problem of predicting ratings for items that have not yet been rated by a target user in a ratings matrix.

As described in Section~\ref{sec:dca}, the data collaboration analysis is aimed at regression and classification tasks, where a single response variable is predicted from multiple predictor variables. 
However, these tasks are fundamentally different from the rating prediction problem, which aims at imputing missing values in a rating matrix. 
Therefore, the data collaboration analysis cannot be directly applied to the rating prediction problem for recommender systems. 

\begin{table}[tbh]
    \centering
    \caption{User--item rating matrix}
    \label{tab:matrixbefore}
    \begin{tabular}{c|c|cccc|}
    \multicolumn{2}{c}{} & \multicolumn{4}{c}{item} \\ \cline{2-6}
    & & $i_1$ & $i_2$ & $i_3$ & $i_4$ \\ \cline{2-6}
    \multirow{4}{*}{\rotatebox{90}{user~}} 
    & $u_1$ &   & 2 & 1&  \\ 
    & $u_2$ & 3 & 1 & 4 & \\
    & $u_3$ &  & 2 & 3 & 1\\
    & $u_4$ & 3 &  & 4 & 3\\ \cline{2-6}
    \end{tabular}
\end{table}

\begin{table}[tbh]
\centering
\caption{Flattened data format derived from Table 1}
\label{tab:matrixafter}
\scalebox{0.95}{
\begin{tabular}{|c|cccc|cccc|c|}
\multicolumn{1}{c}{} & \multicolumn{4}{c}{user} & \multicolumn{4}{c}{item} \\ \hline
& $u_1$ & $u_2$ & $u_3$ & $u_4$ & $i_1$
& $i_2$ & $i_3$ & $i_4$ & response \\ \hline
1& 1 & 0 & 0 & 0 & 0 & 1 & 0 & 0 & 2 \\ 
2& 1 & 0 & 0 & 0 & 0 & 0 & 1 & 0 & 1 \\
3& 0 & 1 & 0 & 0 & 1 & 0 & 0 & 0 & 3 \\
\vdots & \multicolumn{4}{c|}{\vdots} &
\multicolumn{4}{c|}{\vdots} & \vdots \\ 
10& 0 & 0 & 0 & 1 & 0 & 0 & 1 & 0 & 4 \\ 
11& 0 & 0 & 0 & 1 & 0 & 0 & 0 & 1 & 3 \\ \hline
\end{tabular}
}
\end{table}

To overcome the aforementioned challenge, we focus on the factorization machines~\citep{rendle2010factorization} as a rating prediction algorithm. 
This method treats the rating prediction problem as a regression task by converting the rating matrix into a flattened data format as shown in Table~\ref{tab:matrixafter}.
Here, dummy variables for users and items are created as predictor variables, and the corresponding ratings are employed as a response variable. 
Additionally, other attributes such as user gender and item category can readily be added as predictor variables in factorization machines. 
This data format conversion makes it possible to apply the data collaboration analysis to recommender systems. 

\subsection{Our algorithm} \label{sec:proposed} 
Algorithm~\ref{alg:proposed} describes our rating prediction algorithm using the data collaboration analysis. 
We suppose that each party $k \in [m]$ possesses its user--item rating matrix $\bm{R}^{(k)} \in \mathbb{R}^{|U(k)| \times |I|}$, where $U(k)$ is the user index set in party $k \in [m]$, and $I$ is the common item index set. 


\begin{algorithm}[tb]
    \small
    \caption{Rating prediction using the data collaboration analysis}
    \label{alg:proposed}
    \begin{algorithmic}[1]
    \REQUIRE $\bm{R}^{(k)} \in \mathbb{R}^{|U(k)| \times |I|} \quad (k \in [m])$.
    \ENSURE $h \circ g_k \circ f_k:~\mathbb{R}^{1 \times p} \to \mathbb{R} \quad (k \in [m])$. \\
    \COMMENT{Phase 0 (party-side)$.$ Data preparation}
    \STATE Convert $\bm{R}^{(k)}$ into $(\bm{X}^{(k)},\bm{y}^{(k)})$  for each $k \in [m]$. 
    \STATE Share $\bm{S}$ among all parties. \\
    \COMMENT{Phase 1 (party-side)$.$ Construction of intermediate representations}
    \STATE Apply $f_k$ to $\bm{X}^{(k)}$ and $\bm{S}$ to obtain $\tilde{\bm{X}}^{(k)}$ and $\tilde{\bm{S}}^{(k)}$ for each $k \in [m]$. 
    \STATE Send $\tilde{\bm{X}}^{(k)}$, $\tilde{\bm{S}}^{(k)}$, and $\bm{y}^{(k)}$ for all $k \in [m]$ to the analyzer.\\
    \COMMENT{Phase 2 (analyzer-side)$.$ Construction of collaboration representations}
    \STATE Compute the singular value decomposition of $\bm{W}_{\tilde{S}}$ to obtain $\bm{Z}$.  
    \STATE Calculate $\bm{G}^{(k)} = (\tilde{\bm{S}}^{(k)})^\dagger \bm{Z}$ as the solution to problem (Eq.~\eqref{obj:prob0}) for $k \in [m]$. 
    \STATE Calculate $\hat{\bm{X}}^{(k)} = \tilde{\bm{X}}^{(k)} \bm{G}^{(k)}$ for $k \in [m]$.
    \STATE Merge $\hat{\bm{X}}^{(k)}$ and $\bm{y}^{(k)}$ for all $k \in [m]$ to obtain $(\hat{\bm{X}},\bm{y})$ as in Eq.~\eqref{eq:data_merge}. \\
    \COMMENT{Phase 3 (analyzer-side)$.$ Collaborative rating prediction}
    \STATE Train $h$ from $(\hat{\bm{X}},\bm{y})$.
    \STATE Return $h$ and $g_{k}$ (i.e., $\bm{G}^{(k)}$) to each party $k \in [m]$. 
    \end{algorithmic}
\end{algorithm}

Each party $k \in [m]$ first converts the data format of rating matrix from $\bm{R}^{(k)} \in \mathbb{R}^{|U(k)| \times |I|}$ to $(\bm{X}^{(k)},\bm{y}^{(k)}) \in \mathbb{R}^{n(k) \times p} \times \mathbb{R}^{n(k) \times 1}$ (e.g., from Table~\ref{tab:matrixbefore} to Table~\ref{tab:matrixafter}), as defined by Eq.~\eqref{eq:data_party}. 
Here, $n(k)$ is the number of ratings held by party $k \in [m]$, and $p \coloneqq \sum_{k=1}^{m}|U(k)| + |I|$ is the number of dummy variables corresponding to users and items. 
All parties also share an anchor dataset $\bm{S} \in \mathbb{R}^{r \times p}$ for data collaboration. 

In the first phase, each party $k \in [m]$ applies its encoding function (Eq.~\eqref{eq:enc_fun}) to the predictor matrix $\bm{X}^{(k)}$ and the anchor dataset $\bm{S}$, thereby yielding the privacy-preserving intermediate representations (Eq.~\eqref{eq:data_intrep}). 
Then, all parties $k \in [m]$ send $\tilde{\bm{X}}^{(k)}$, $\tilde{\bm{S}}^{(k)}$, and $\bm{y}^{(k)}$ to the analyzer. 

In the second phase, the analyzer calculates the collaboration representations (Eq.~\eqref{eq:data_clbrep}) based on the integration functions (Eq.~\eqref{eq:lin_int}) according to the procedure described in Section~\ref{sec:cr}. 
After that, the analyzer merges the resultant datasets as 
\begin{align}
\hat{\bm{X}} \coloneqq \begin{pmatrix}
\hat{\bm{X}}^{(1)} \\
\hat{\bm{X}}^{(2)} \\
\vdots \\
\hat{\bm{X}}^{(m)}
\end{pmatrix} \in \mathbb{R}^{n \times \hat{p}}, \quad 
\bm{y} \coloneqq \begin{pmatrix}
\bm{y}^{(1)} \\
\bm{y}^{(2)} \\
\vdots \\
\bm{y}^{(m)}
\end{pmatrix} \in \mathbb{R}^{n \times 1}. \label{eq:data_merge}
\end{align}

In the third phase, the analyzer uses the merged dataset $(\hat{\bm{X}},\bm{y})$ to train a machine learning model (Eq.~\eqref{eq:ML_model}) for rating prediction. 
The analyzer then returns the trained model $h$ and integration function $g_k$ to each party $k \in [m]$. 
As a result, each party $k \in [m]$ obtains a highly accurate model (Eq.~\eqref{eq:fin_model}) for rating prediction. 

Note that Algorithm~\ref{alg:proposed} is aimed at the \emph{horizontal partitioning}~\citep{yang2020federated}, where items are shared but users are different between parties; however, this algorithm can be readily applicable to the \emph{vertical partitioning}~\citep{yang2020federated}, where users are shared but items are different among parties. 
Specifically, the same algorithm can be used after transposing rating matrices. 

\section{Numerical experiments}
In this section, we evaluate the effectiveness of our privacy-preserving method for rating prediction through numerical experiments. 

\subsection{Experimental setup}
We used two public rating datasets, namely the MovieLens 100K\footnote[1]{\url{https://grouplens.org/datasets/movielens/100k/}} and SUSHI preference\footnote[2]{\url{https://www.kamishima.net/sushi/}} datasets. 
The MovieLens 100K dataset contains 100,000 ratings for 1,682 movies from 943 users on a scale of 1 to 5. 
The SUSHI preference dataset contains ratings for 100 sushi items from 5,000 users, where each user rated 10 sushi items on a scale of 0 to 4. 
For reference, Figure~\ref{fig:rate} shows the frequency distributions of ratings in these datasets. 

Ratings of each user were randomly split into training (80\%) and testing (20\%) datasets. 
These datasets were randomly distributed to 9 and 50 parties for the MovieLens 100K and SUSHI preference datasets, respectively, with each party holding ratings from 100 users. 
Machine learning models were trained on the training dataset, and prediction accuracy was measured by the root mean squared error (RMSE) in user ratings on the testing dataset. 
The process of dataset generation and accuracy evaluation was repeated 10 times, and the average RMSE values with standard errors are given as numerical results.

\begin{figure}[tb]
\centering
\begin{subfigure}[t]{0.495\textwidth}
\centering
\includegraphics[scale=0.24]{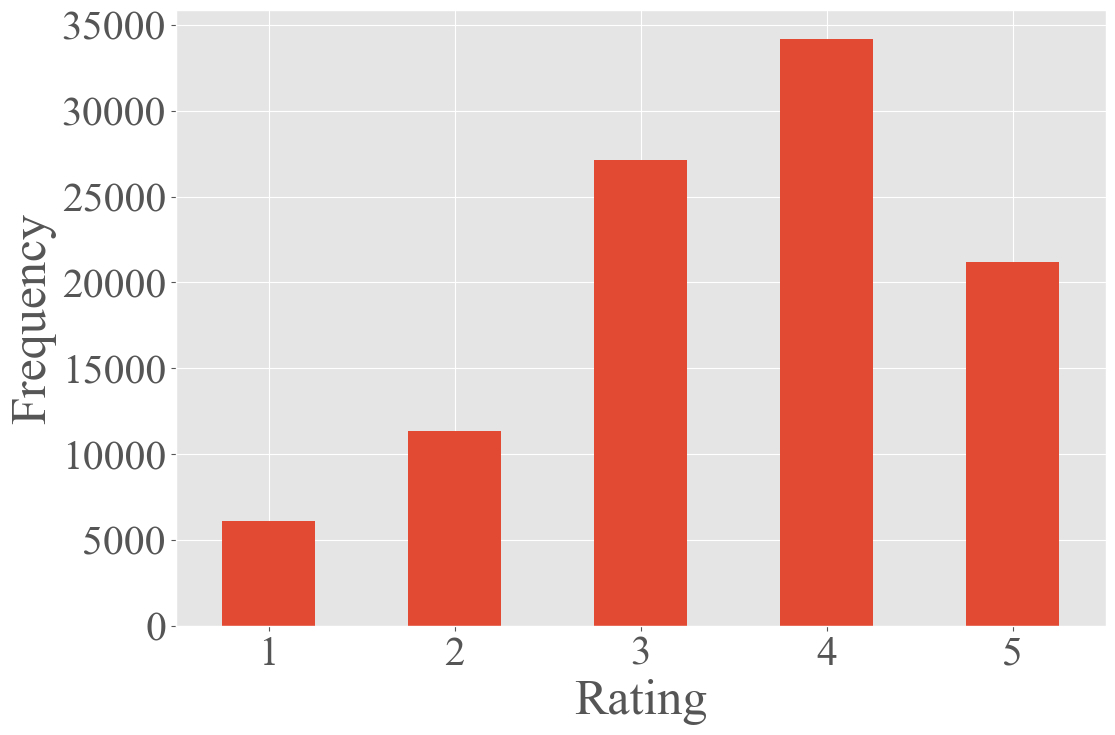}
\caption{MovieLens 100K dataset}\label{fig:movierate}
\end{subfigure}
\begin{subfigure}[t]{0.495\textwidth}
\centering
\includegraphics[scale=0.24]{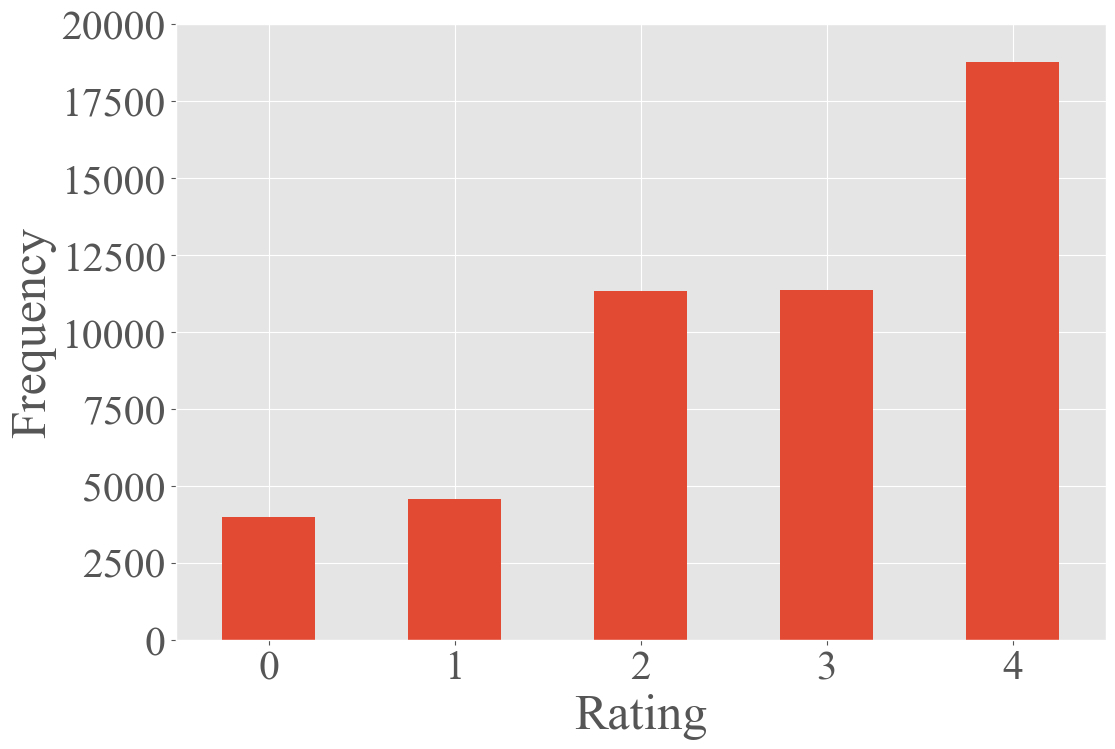}
\caption{SUSHI preference dataset}\label{fig:sushirate}
\end{subfigure}
\caption{Frequency distributions of ratings}
\label{fig:rate}
\end{figure}

We compare the prediction accuracy of the following methods for analyzing distributed datasets (Figure~\ref{fig:rate}): 
\begin{description}
    \item[Individual analysis:] individual machine learning models are trained by each party using only its own dataset (Eq.~\eqref{eq:data_party});
    \item[Centralized analysis:] a common machine learning model is trained by all parties using the merged dataset (Eq.~\eqref{eq:data_all}), while privacy concerns are disregarded;  
    \item[Data collaboration analysis:] a common machine learning model is trained using the data collaboration analysis for privacy protection (Algorithm~\ref{alg:proposed}). 
\end{description} 

\begin{figure}[tb]
\centering
\begin{subfigure}[t]{0.32\textwidth}
\centering
\includegraphics[scale=0.09]{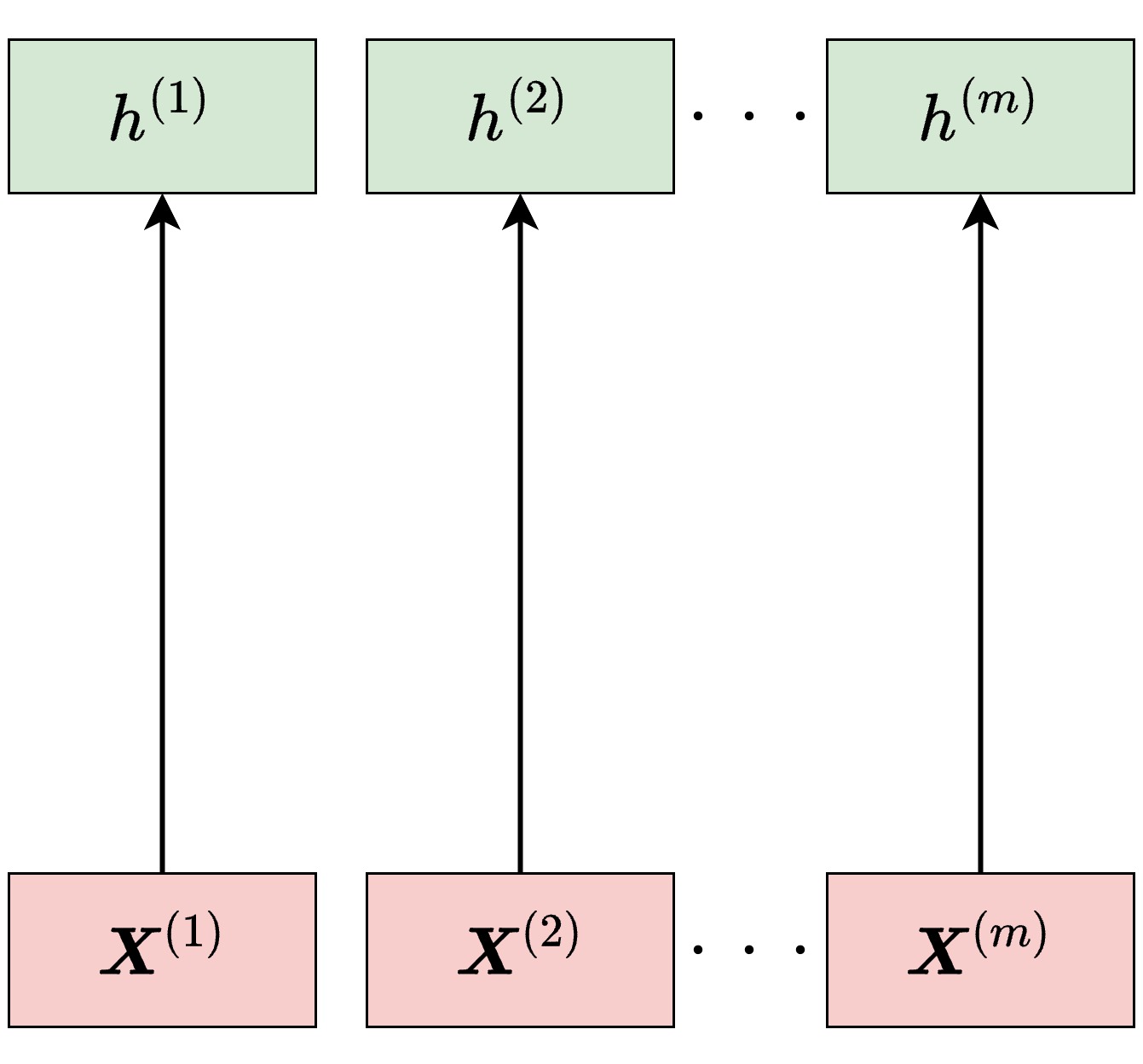}
\caption{Individual analysis}\label{fig:indiv_analysis}
\end{subfigure}
\begin{subfigure}[t]{0.32\textwidth}
\centering
\includegraphics[scale=0.09]{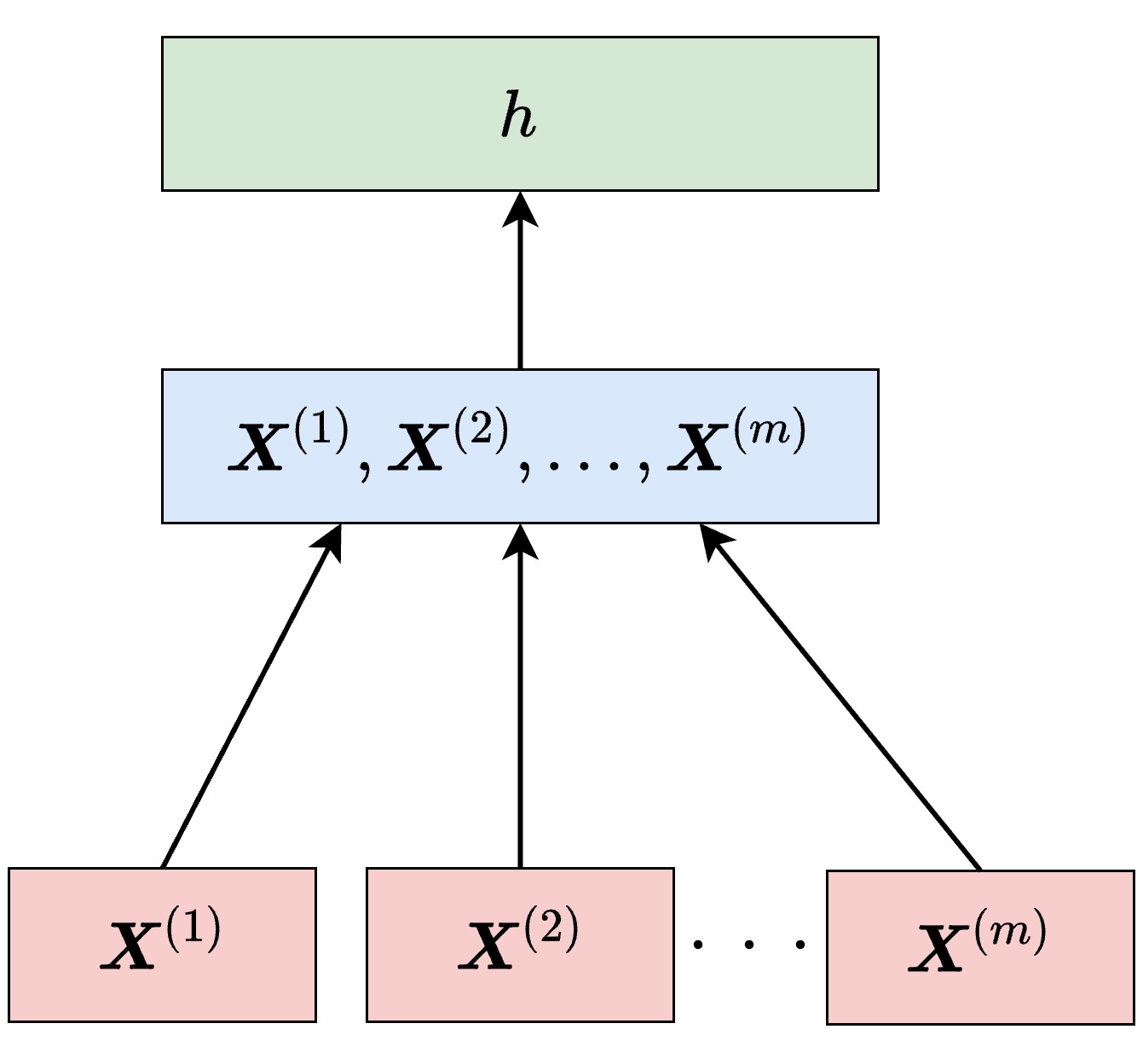}
\caption{Centralized analysis}\label{fig:center_analysis}
\end{subfigure}
\begin{subfigure}[t]{0.32\textwidth}
\centering
\includegraphics[scale=0.09]{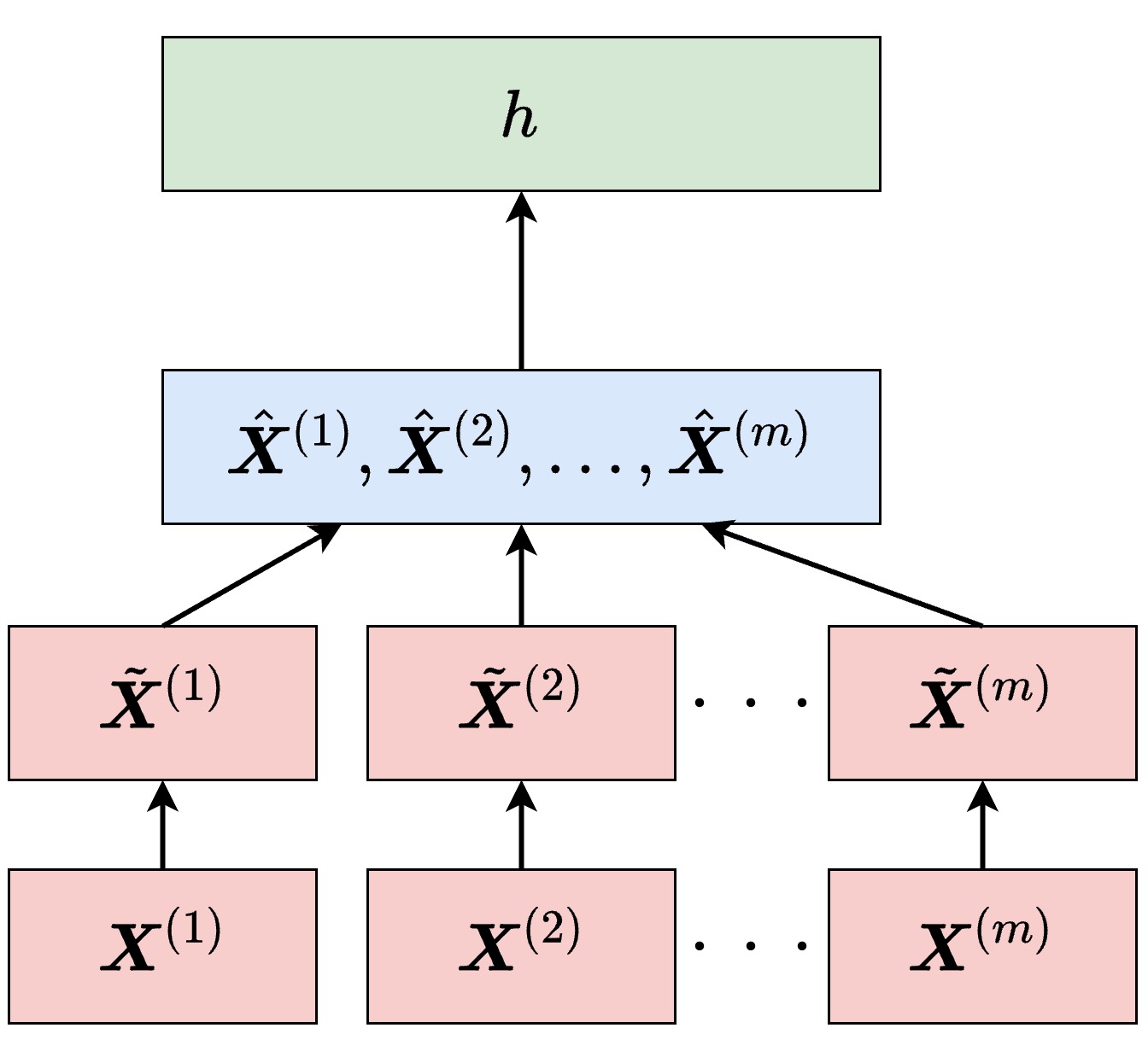}
\caption{Data collaboration analysis}\label{fig:collab_analysis}
\end{subfigure}
\caption{Workflow of the methods for analyzing distributed datasets}
\label{fig:rate}
\end{figure}


For the data collaboration analysis, an anchor dataset (Eq.~\eqref{eq:data_anchor}) containing 1,000 instances (i.e., $r=1{,}000$) was generated from a uniform distribution over the interval $[0,1]$. 
The intermediate representations (Eq.~\eqref{eq:data_intrep}) were created using the singular value decomposition for dimensionality reduction. 
The numbers of dimensions for the intermediate and collaboration representations were set as $\tilde{p}(k) \in \{100, 200, 400\}$ and $\hat{p} \in \{\tilde{p}^{(k)}/2, \tilde{p}^{(k)}, 2\tilde{p}^{(k)}\}$, respectively for $k \in [m]$. 
As in the data collaboration analysis, the flattened data format $(\bm{X}^{(k)},\bm{y}^{(k)})$ (e.g., Table~\ref{tab:matrixafter}) was used to train machine learning models in both the individual and centralized analyses. 

For rating prediction, we used the following machine learning models: 
\begin{description}
    \item[pyFM:] a Python implementation of factorization machines\footnote[3]{\url{https://github.com/coreylynch/pyFM}};
    \item[LightGBM:] a framework of gradient boosting decision trees~\citep{ke2017lightgbm}. 
\end{description}
The length of latent vectors in the pyFM was set to 100. 
To mitigate the sparsity of datasets, the number of dataset dimensions for the LightGBM was reduced to 200 through the singular value decomposition. 
The hyperparameters of the LightGBM were tuned through 5-fold cross-validation using Optuna~\citep{akiba2019optuna}, a library for Bayesian optimization. 

\subsection{Numerical results}
Tables~\ref{tab:result_movie} and \ref{tab:result_sushi} give the testing RMSEs provided by the rating prediction methods for the MovieLens 100K and SUSHI preference datasets, respectively. 
Recall that these tables give average values over 10 repetitions, with standard errors in parentheses. 


\begin{table}[tb]
    \centering
    \caption{RMSE for the  MovieLens 100K dataset ($m = 9$)}
    \label{tab:result_movie}
    \begin{tabular}{crrrr} \toprule
    &&&\multicolumn{2}{c}{Machine learning model} \\ \cmidrule(lr){4-5} 
    Analysis method & $\tilde{p}^{(k)}$ & $\hat{p}$ & \multicolumn{1}{c}{pyFM} & \multicolumn{1}{c}{LightGBM} \\ \midrule
    Individual analysis & --- & --- & \textbf{1.072}~($\pm 0.007$) & \textbf{1.080}~($\pm 0.009$) \\ \midrule
    Centralized analysis & --- & --- & \textbf{0.948}~($\pm 0.002$) & \textbf{0.978}~($\pm 0.002$) \\ \midrule
    Data collaboration & 100 & 50 & 1.063~($\pm 0.003$) & 1.027~($\pm 0.002$) \\
     analysis & & 100 & 1.058~($\pm 0.004$) & 1.024~($\pm 0.001$) \\
     & & 200 & \textbf{1.052}~($\pm 0.002$) & \textbf{1.021}~($\pm 0.002$) \\ \cmidrule{2-5}
     & 200 & 100 & 1.063~($\pm 0.002$) & 1.021~($\pm 0.003$) \\
     & & 200 & 1.055~($\pm 0.002$) & \textbf{1.012}~($\pm 0.002$) \\
     & & 400 & \textbf{1.048}~($\pm 0.002$) & 1.014~($\pm 0.001$) \\ \cmidrule{2-5}
     & 400 & 200 & 1.060~($\pm 0.002$) & 1.020~($\pm 0.002$) \\
     & & 400 & 1.055~($\pm 0.002$) & 1.012~($\pm 0.002$) \\
     & & 800 & \textbf{1.048}~($\pm 0.001$) & \textbf{1.010}~($\pm 0.001$) \\ \bottomrule
    \end{tabular}
\end{table}

\begin{table}[tb]
    \centering
    \caption{RMSE for the  SUSHI preference dataset ($m = 50$)}
    \label{tab:result_sushi}
    \begin{tabular}{crrrr} \toprule
    &&&\multicolumn{2}{c}{Machine learning model} \\ \cmidrule(lr){4-5}
    Analysis method & $\tilde{p}^{(k)}$ & $\hat{p}$ & \multicolumn{1}{c}{pyFM} & \multicolumn{1}{c}{LightGBM} \\ \midrule
    Individual analysis & --- & --- & \textbf{1.240}~($\pm 0.009$) & \textbf{1.257}~($\pm 0.009$) \\ \midrule
    Centralized analysis & --- & --- & \textbf{1.134}~($\pm 0.004$) & \textbf{1.145}~($\pm 0.006$) \\ \midrule
    Data collaboration & 100 & 50 & 1.221~($\pm 0.002$) & 1.197~($\pm 0.003$) \\
    analysis & & 100 & 1.219~($\pm 0.003$) & 1.194~($\pm 0.003$) \\
    & & 200 & \textbf{1.214}~($\pm 0.002$) & \textbf{1.177}~($\pm 0.003$) \\ \cmidrule{2-5}
    & 200 & 100 & 1.215~($\pm 0.003$) & 1.195~($\pm 0.003$) \\
    & & 200 & 1.216~($\pm 0.003$) & 1.188~($\pm 0.002$) \\
    & & 400 & \textbf{1.209}~($\pm 0.003$) & \textbf{1.172}~($\pm 0.002$) \\ \cmidrule{2-5}
     & 400 & 200 & 1.215~($\pm 0.003$) & 1.195~($\pm 0.001$) \\
     & & 400 & 1.214~($\pm 0.002$) & 1.187~($\pm 0.001$) \\
     & & 800 & \textbf{1.208}~($\pm 0.001$) & \textbf{1.169}~($\pm 0.002$) \\ \bottomrule
    \end{tabular}
\end{table}

First, we compare the performance of the three analysis methods (i.e., individual, centralized, and data collaboration analyses). 
For both datasets, the RMSEs of the data collaboration analysis were larger than those of the centralized analysis and smaller than those of the individual analysis. 
Recall that the individual analysis does not merge distributed datasets, whereas the centralized analysis merges distributed datasets without protecting privacy. 
For these reasons, the obtained results are considered reasonable.

Next, we discuss the dimensionality of intermediate and collaboration representations (i.e., $\tilde{p}^{(k)}$ and $\hat{p}$) employed in the data collaboration analysis. 
As for the dimensionality of intermediate representations, the lowest RMSEs were obtained when setting $\tilde{p}^{(k)}=400$. 
This is because low-dimensional intermediate representations lose information from the original datasets. 
As for the dimensionlity of collaboration representations, the lowest RMSEs were obtained when setting $\hat{p} = \tilde{p}^{(k)}$ or $\hat{p} = 2 \tilde{p}^{(k)}$. 
This is probably because high-dimensional collaboration representations increase the flexibility of integration functions. 

Then, we focus on the machine learning models (i.e., pyFM and LightGBM) implemented for rating prediction. 
The factorization machines (pyFM) consistently outperformed the gradient boosting decision trees (LightGBM) for the individual and centralized analyses, whereas the opposite results were obtained for the data collaboration analysis. 
Although factorization machines generally perform well for rating prediction, the data collaboration analysis involves dimensionality reduction for creating intermediate representations, thereby improving prediction performance of gradient boosting decision trees. 

\begin{figure}[tb]
\centering
\begin{subfigure}[t]{0.495\textwidth}
\centering
\includegraphics[scale=0.2]{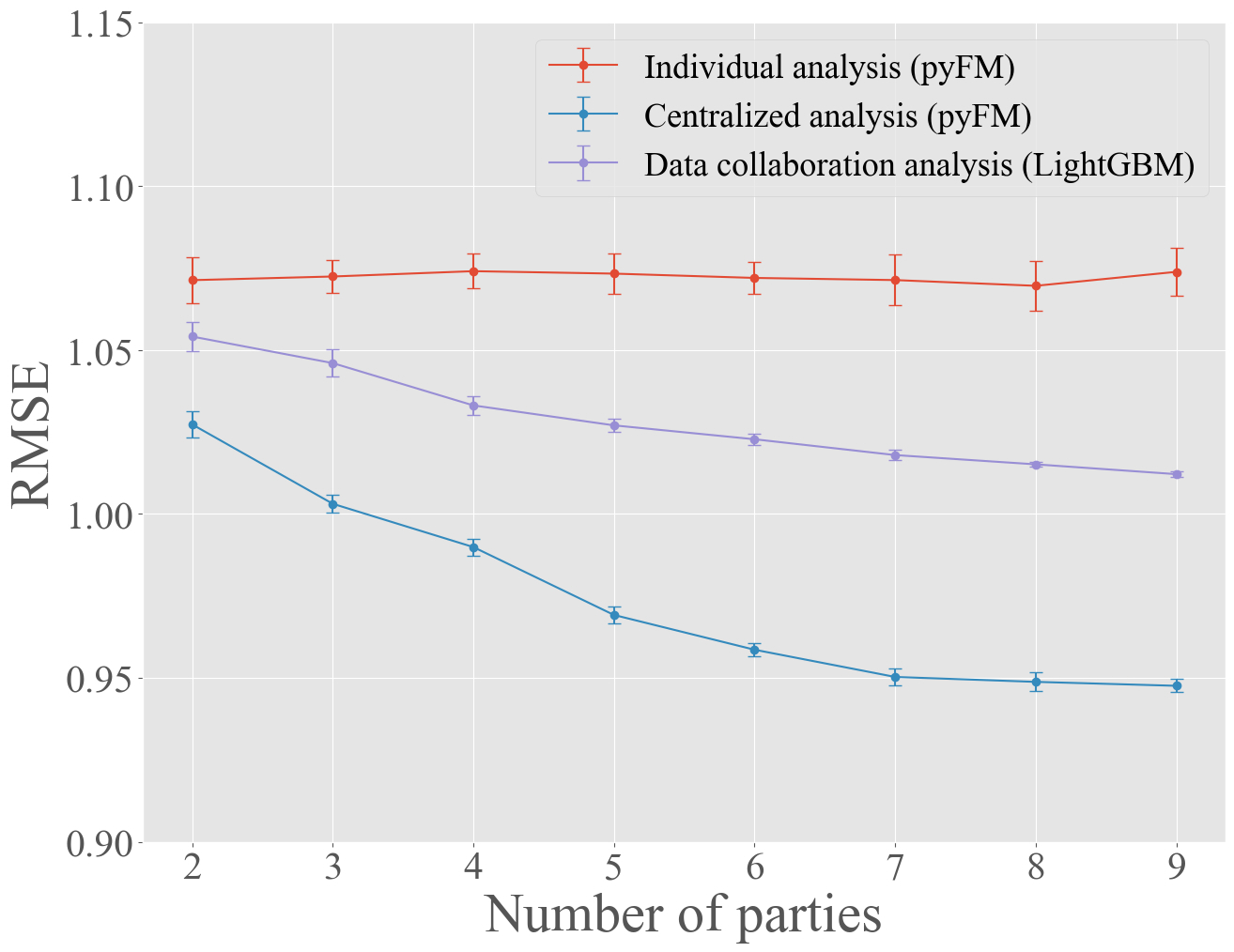}
\caption{MovieLens 100K dataset}\label{fig:result_movie}
\end{subfigure}
\begin{subfigure}[t]{0.495\textwidth}
\centering
\includegraphics[scale=0.2]{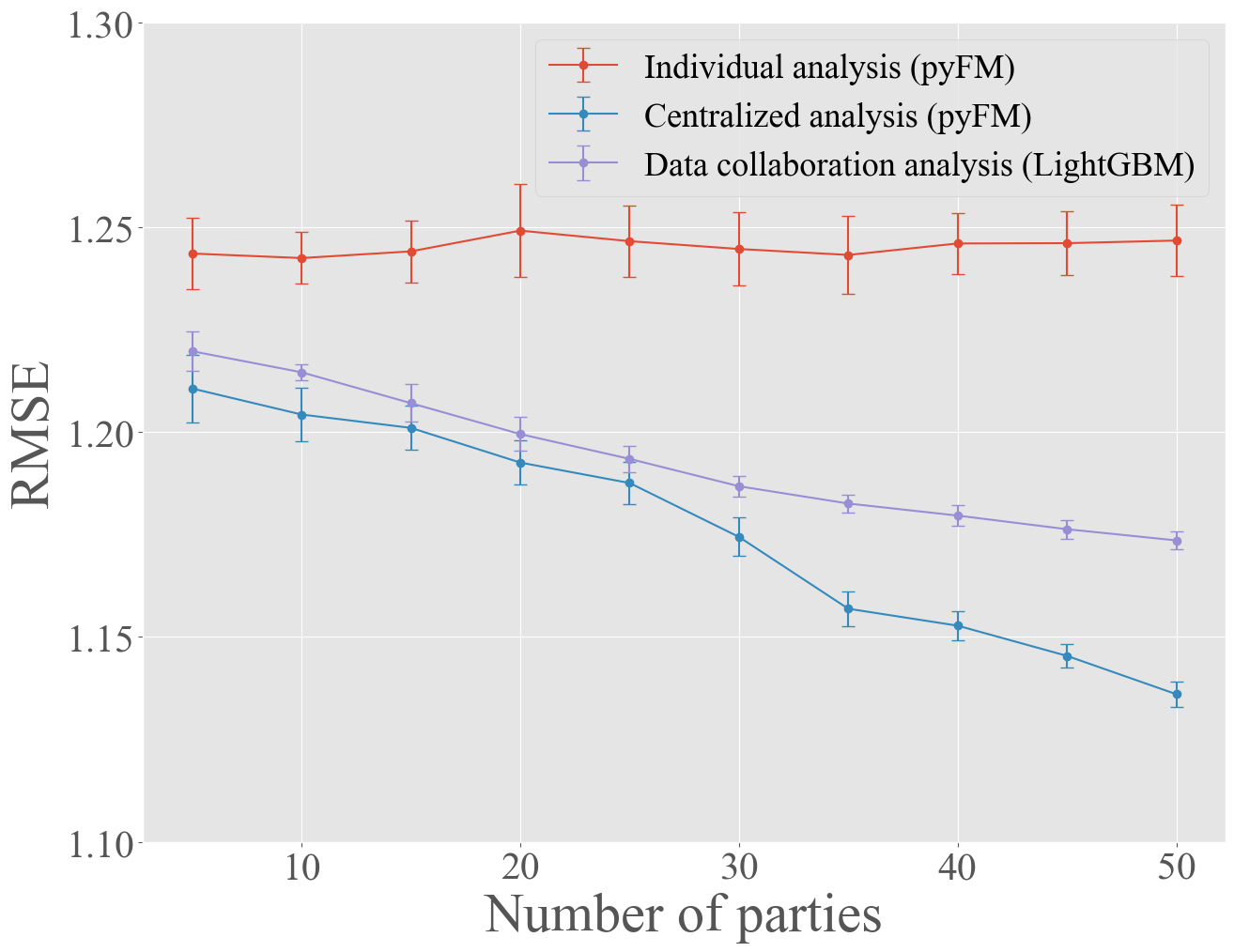}
\caption{SUSHI preference dataset}\label{fig:result_sushi}
\end{subfigure}
\caption{RMSE as a function of the number of involved parties}
\label{fig:result}
\end{figure}

Figure~\ref{fig:result} shows the testing RMSEs provided by the rating prediction methods as a function of the number of involved parties for the MovieLens 100K and SUSHI preference datasets. 
Recall that this figure shows average values over 10 repetitions, with standard errors shown as error bars. 
Note here that the factorization machines were used for the individual and centralized analyses, and that the gradient boosting decision trees were used for the data collaboration analysis; this is due to the effectiveness of these combinations, as previously discussed. 
The numbers of dimensions for the intermediate and collaboration representations were set as $\tilde{p}(k) = 200$ and $\hat{p} = 400$ for $k \in [m]$. 

Naturally, increasing the number of parties did not improve prediction accuracy in the individual analysis. 
In contrast, the RMSEs decreased with the increasing number of parties in the centralized and data collaboration analyses. 
Additionally, although the individual analysis always showed relatively large standard errors of RMSEs, the data collaboration analysis decreased the standard errors as the number of parties increased. 
These results demonstrate that increasing the number of involved parties yields more accurate and stable prediction models in the data collaboration analysis. 

\section{Conclusion}
We focused on privacy-preserving recommender systems on distributed datasets. 
For this purpose, we designed a rating prediction algorithm using the data collaboration analysis~\citep{imakura2020data} for privacy protection. 
In this algorithm, the user--item rating matrix is converted into the flattened format with the aim of treating missing value imputation as regression. 
This conversion makes it possible to apply the data collaboration analysis to rating prediction for recommendations. 
Note also that our algorithm is readily applicable to both horizontal and vertical integration of rating matrices.

To verify the effectiveness of our method, we performed numerical experiments using two public rating datasets. 
Our method for collaborative rating prediction improved the prediction accuracy while protecting privacy of the original datasets. 
Even when each party owns a small dataset, our method can produce a reliable recommendation system through data collaboration.
Numerical results also confirmed that the prediction accuracy of our method surpassed that of the individual analysis for both datasets and was comparable to that of the centralized analysis, particularly in the SUSHI preference dataset.

A future research direction will be to use more sophisticated methods for creating anchor datasets \citep{takahashi2021decentralized,imakura2023another} and integration functions~\citep{kawakami2024new,nosaka2023creating} in the data collaboration analysis. 
Other directions include the application of the data collaboration analysis to generating a high-quality list of recommendations~\citep{wang2009mean,hurley2011novelty,yasumotoamean} and promoting sales through price optimization~\citep{klein2020review,ikeda2023prescriptive}.

\section*{Acknowledgments}
This work was supported by JSPS KAKENHI Grant Number JP21K04526. 
\bibliographystyle{elsarticle-harv} 
\bibliography{cite.bib}





\end{document}